\documentclass[12pt]{article}
\usepackage[english]{babel}
\usepackage{graphicx,graphics}
\newcommand {\vT} {v _ {\scriptscriptstyle T}}
\newcommand {\di} {\displaystyle}
\newcommand {\p} {\partial}

\begin{document}
\title{Piecewise continuous distribution function method and ultrasound at half space}
\author{ Solovchuk M. A., Leble S. B.\\
\small Theoretical Physics Department,
\small Immanuel Kant State
University of Russia, Russia, \\
\small 236041, Kaliningrad, Al. Nevsky str. 14. \\
\small   Theoretical Physics and Mathematical Methods Department,\\
\small  Technical University of Gdansk, ul, Narutowicza 11/12,
Gdansk,  Poland,\\
\small  leble@mifgate.pg.gda.pl \\  \\[2ex] }
\maketitle

\begin{abstract}
 The system of hydrodynamic-type equations, derived by two-side distribution function
  for a stratified gas in
gravity field is applied to a problem of ultrasound propagation
and attenuation. The background state and linearized version of
the obtained system is studied and compared with the Navier-Stokes
one at arbitrary Knudsen numbers. The WKB solutions for ultrasound
in a stratified medium are constructed in explicit form. The
problem of a generation by a moving plane in a rarefied gas is
explored and used as a test while compared with experiment.
\end{abstract}

\section{Introduction}

Recently  the problems of Kn regime wave propagation was revisited
in connection with general fluid mechanics and nonsingular
perturbation method development \cite{veresc3,
Chen2000,Chen2001b,spiegel}. A generalized Boltzman theories
\cite{alexeev2004,EChe} also contributed in a progress with
respect to this important problem.

In \cite{LSV} the propagation of one-dimension disturbance was
studied on the base of  the method of a piecewise continuous
distribution function launched in a pioneering paper of Lees
\cite{lees} and applied for a gas in gravity field in
\cite{veresc3,VerLe}. We derived hydrodynamic-type equations for a
gas perturbations in gravity field so that the Knudsen number
depends on the (vertical) coordinate. The generalization to three
dimensions is given at \cite{Budapest2005}

The derivation of the hydrodynamic-type equations is based on
kinetic equation with the model integral of collisions in BGK
(Bhatnagar - Gross - Krook ) form which collision term is modelled
as $\nu\left(f_{\it l}-f\right)\ ,$ via local-equilibrium
distribution function $f_{\it l}$ and the non-equilibrium one is
expressed as $f^+$ at $v_z\geq 0$, and as $f^-$ at $v_z \leq 0$
$$
f^{\pm}=\di \frac{n^{\pm}}{\pi^{3/2}\vT^{\pm 3}} \exp(-\frac{(\vec
V-\vec U^\pm)^2}{\vT^{\pm 2}}),
$$
the  $ \vT =\sqrt{2kT/m} $ denotes the average thermal velocity of
particles of gas, $ \nu =\nu(z) $ is the effective frequency of
collisions between particles of gas at height $z $. It is
supposed, that density of gas $n $, its average speed $ \vec U =
(u_x, u_y, u_z) $ and temperature $T $ are functions of time and
coordinates enter the local-equilibrium $f_{\i l}$. The resulting
system is

\begin{equation}
\label{q6}
\begin{array}{l}
\di
 \frac{\p}{\p t}\rho + \frac{\p}{\p z}(\rho U) = 0
\ ,\vspace{2mm} \\
 \di
 \frac{\p}{\p t}U + U \frac{\p}{\p z}U +
 \frac{1}{\rho} \frac{\p}{\p z}P_{zz} + g = 0
\ ,\vspace{2mm} \\
 \di
 \frac 32 \frac {k}{m} \frac{\p}{\p t}( \rho T) + \frac 32 \frac {k}{m}U \frac{\p}{\p z}( \rho T) +
 (\frac 32 \frac {k}{m} \rho T + P_{zz})\frac{\p}{\p z}U + \frac{\p}{\p z}q_z = 0
\ ,\vspace{2mm} \\
 \di
 \frac{\p}{\p t}P_{zz} + U \frac{\p}{\p z}P_{zz} +
 3 P_{zz}\frac{\p}{\p z}U + 2 \frac{\p}{\p z}\bar q_z =
 - \nu(z) (P_{zz} -  \frac {\rho}{m} k T )
\ ,\vspace{2mm} \\
  \di
  \frac{\p}{\p t}q_z + U \frac{\p}{\p z}q_z +
 2 (q_z + \bar q_z)\frac{\p}{\p z}U - (\frac 32 \frac {k}{m} T + \frac{1}{\rho} P_{zz})
\frac{\p}{\p z}P_{zz} +
 \frac{\p}{\p z}J_1 = - \nu(z) q_z
\ ,\vspace{2mm} \\
 \di
 \frac{\p}{\p t}\bar q_z + U \frac{\p}{\p z}\bar q_z +
 4 \bar q_z \frac{\p}{\p z}U -
 \frac{3}{2\rho} P_{zz} \frac{\p}{\p z}P_{zz} +
 \frac{\p}{\p z}J_2 = - \nu(z) \bar q_z\ ,
\end{array}
\end{equation}
where
\begin{equation}
\label{q2}
\begin{array}{ll}
 \di  \quad  J_1 = \frac m2 <(V_z - U)^2 (\vec V - \vec  U)^2> \ ,
 \quad &
\di J_2 = \frac m2 <(V_z - U)^4>\ . \\
\end{array}
\end{equation}

 The increase of the number of parameters of
distribution function results in that the distribution function
differs from a local-equilibrium one and describes deviations from
hydrodynamical regime. In the range of small Knudsen numbers $
{\it l}<< L $ we automatically have $ n^+ = n^-, U^+ = U^-, T^+ =
T^-$ and distribution function reproduces the hydrodynamics of
Euler and at the small difference of the functional "up" and
"down" parameters  - the Navier-Stokes equations. In the range of
big Knudsen numbers the theory gives solutions of collisionless
problems \cite{VerLe}.

We used a set of linearly independent eigen functions of the
linearized Boltzmann operator, that in the case of the BGK
equation is:
 \begin{equation}
\label{q3}
\begin{array}{rclrrcl}
 \varphi_1&=&m\ ,
 & \varphi_4&=& m(V_z - U_z)^2\ , \\
 \varphi_2&=&m V_z\ ,
 & \varphi_5&=&\di \frac{1}{2} m(V_z-U_z)|\vec V - \vec U|^2\ ,\\
 \varphi_3&=&\di \frac{1}{2}m |\vec V - \vec U|^2\ ,
 & \varphi_6&=&\di \frac{1}{2} m(V_z-U_z)^3\ .\\
\end{array}
\end{equation}

Let's define a scalar product in velocity space:
\begin{equation}
\label{q4}
 <\varphi_n,f> \equiv <\varphi_n>
 \equiv \int d\vec v\: \varphi_n(t,z,\vec V) f(t,z,\vec V)\ .
\end{equation}

\begin{equation}
\label{q5}
\begin{array}{lll}
 <\varphi_1> = \rho(t, z) \ ,
  &  <\varphi_2>= \rho U \ ,
 & <\varphi_3> =\frac 32 \frac {\rho}{m}k T \ , \\
<\varphi_4> =  P_{zz}\ , & <\varphi_5>  = q_z, \
 & <\varphi_6> = \di  \bar q_z .\\
\end{array}
\end{equation}

Here $\rho$ is mass density, $P_{zz}$ is the diagonal component of
the pressure tensor, $q_z$ is a vertical component of a heat flow,
$\bar q_z$ is a parameter having dimension of the heat flow.

The system (\ref{q6}) of the equations according to the derivation
scheme is valid at all frequencies of collisions and within the
limits of the high frequencies should transform to the
hydrodynamic equations.

If we estimate the functions  $ \frac {U ^ {\pm}} {V_T ^ {\pm}}  $
as small, that corresponds to small Mach numbers $M = max| \di
\frac {U}{v_T}| \ $.  We shall base here on an expansion in $ M$,
up to the first order. In this approach the functional parameters
of the two-fold distribution function

$$
\begin{array}{lll}
n^+=n_0(1+\alpha n^+_1) \ , & n^-=n_0(1+\alpha n^-_1) \ , & \rho=nm \  \vspace{1mm}  \\
V^+=V_0(1+\alpha V^+_1) \ , & V^-=V_0 (1+\alpha V^-_1)\ , \vspace{1mm} \\
U^+=\alpha V_0 U^+_1\ , \quad & U^-=\alpha V_0 U^-_1 \
\end{array}
$$

Let's evaluate  the integrals (\ref {q2}) and (\ref {q5})
directly, plugging the two-side distribution function. In the
first order by Mach number $\alpha$

\begin{equation}
\label{q7}
\begin{array}{l}
n=n_0+ \left( {\frac {n_0 U^+_1 }{\sqrt {\pi }}}+ \frac 12 n_0
n^+_1 + \frac 12  n_0 n^-_1 -{\frac {n_0 U^-_1 }{\sqrt {\pi }}}
\right) \alpha
\\
U=\left( - \frac 12  {\frac { V_0 V^-_1 }{\sqrt {\pi }}}- \frac 12
{\frac { V_0n^-_1 }{\sqrt {\pi }}}+ \frac 12  V_0U^+_1 + \frac 12
{\frac {V_0n^+_1 }{\sqrt {\pi }}}+  \frac 12  V_0U^-_1 + \frac 12
{ \frac {V_0 V^+_1 }{\sqrt {\pi }}} \right) \alpha
\\
 \frac 32 \frac {k}{m^2} \rho T= \frac 34  n_0{V_0}^{2}+ \left(
\frac 34 n_0{V_0}^{2} V^+_1 + \frac 38  n_0{V_0}^2 n^+_1 + \frac
38 n_0{V_0}^{2}n^-_1 +{\frac {n_0{V_0}^{2}U^+_1 }{ \sqrt {\pi }}}+
\frac 34  n_0{V_0}^{2} V^-_1 -{\frac {n_0{V_0}^{2}U^-_1 }{\sqrt
{\pi }}} \right) \alpha
\\
\frac 1m P_{zz}= \frac 12  n_0{V_0}^{2}+ \left(  \frac 14
 n_0{V_0}^{2}n^+_1 -{\frac {n_0{V_0}^{2}U^-_1 }{\sqrt {\pi }}} +
\frac 12  n_0{V_0}^{2} V^+_1 + \frac 14  n_0{V_0}^{2 }n^-_1 +
\frac 12  n_0{V_0}^{2} V^-_1 +{\frac {n_0{V_0}^{2}U^+_1 }{\sqrt
{\pi }}} \right) \alpha
\\
\frac 1m q_z=\left( \frac 58 n_0{V_0}^{3}U^+_1 - \frac 54
n_0{V_0}^{2}U + \frac 58 n_0{V_0}^{3}U^-_1 - \frac 32 {\frac
{n_0{V_0}^{3} V^-_1 }{\sqrt {\pi }}}+ \frac 12 {\frac {n_0
{V_0}^{3}n^+_1 }{\sqrt {\pi }}}- \frac 12  {\frac {n_0{V_0
}^{3}n^-_1 }{\sqrt {\pi }}}+ \frac 32  {\frac {n_0{V_0}^{3} V^+_1
}{\sqrt {\pi }}} \right) \alpha
\\
\frac {\bar q_z}{m} =( - \frac 34 n_0{V_0}^{2}U+ \frac 14 {\frac
{n_0{V_0}^{3}n^+_1 }{\sqrt {\pi }}}+ \frac 34 {\frac {n_0{V_0}^
{3} V^+_1 }{\sqrt {\pi }}} - \frac 34 {\frac {n_0{V_0}^{3}
V^-_1}{\sqrt {\pi}}}+ \frac 38  n_0{V_0}^{3}U^-_1 - \frac 14
{\frac {n_0{V_0}^{3}n^-_1 }{\sqrt {\pi }}}+ \frac 38 n_0
{V_0}^{3}U^+_1  ) \alpha .
\end{array}
\end{equation}

\begin{equation}
\label{q8}
\begin{array}{l}
\frac 1m J_1=\frac 58 n_0 {V_0}^{4}+ \left( {\frac {5}{16}}
n_0{V_0 }^{4}n^+_1 +  \frac 54  n_0 {V_0}^{4} V^-_1 - \frac 32
{\frac {n_0{V_0}^{4}U^-_1 }{\sqrt {\pi }}}+ \frac 54
 n_0{V_0}^{4} V^+_1 +{\frac {5}{16}}n_0{V_0}^{ 4}n^-_1 + \frac 32
 {\frac {n_0{V_0}^{4}U^+_1 }{\sqrt {\pi }}} \right) \alpha
\\
\frac 1m J_2=  \frac 38  n_0{V_0}^{4}+ \left(  \frac {3}{16}
 n_0{V_0}^{4}n^+_1 + \frac 34 n_0{V_0}^{4} V^-_1 + \frac {3}{16}
 n_0{V_0 }^{4}n^-_1 -{\frac {n_0{V_0}^{4}U^-_1 }{ \sqrt {\pi}}}+
\frac 34  n_0{V_0}^{4} V^+_1 +{\frac {n_0{V_0}^{4}U^+_1 }{\sqrt
{\pi }}} \right) \alpha
\end{array}
\end{equation}

Solving the system (\ref {q7}), we obtain for the parameters of
the two-fold distribution function:
$$
\begin{array}{l}
\di n_1^+= -\frac 32 {\frac {U \sqrt {\pi}}{V_0}}+{\frac
{n}{n_0}}-1-3 \frac {P_{zz}}{m n_0 {V_0}^{2}} + 3{\frac { k \rho
T}{n_0 m^2{V_0}^{2}}}- 7 \frac { \bar q_z \sqrt {\pi}}{m n_0
{V_0}^3} + 3 {\frac { q_z \sqrt {\pi}}{m n_0{V_0}^3}}+ \frac 32
 {\frac { nU \sqrt {\pi}}{n_0 V_0}} ,\
 \\
\di n_1^-= \frac 32  {\frac {U\sqrt {\pi }}{V_0}}+{\frac
{n}{n_0}}-1-3 { \frac { P_{zz}}{m n_0{V_0}^{2}}} + 3 \frac {k \rho
T}{n_0 m^2{V_0}^2} +7 {\frac { \bar q_z \sqrt {\pi}}{m
n_0{V_0}^3}}-3 {\frac { q_z \sqrt {\pi }}{m n_0{V_0}^3}}- \frac 32
 {\frac {{\it nU} \sqrt {\pi }}{n_0 V_0}} ,\
\\
\di V_1^+= - \frac 12  {\frac {{\it nU} \sqrt {\pi }}{V_0 n_0}}-
\frac 12  \frac {P_{zz}}{m n_0{V_0}^2} + \frac 32 {\frac { k \rho
T}{n_0 m^2{V_0}^2}}- \frac 12  {\frac {n}{n_0}}+{\frac { q_z \sqrt
{\pi}}{mn_0{V_0}^3}} + \frac 12  {\frac {U\sqrt {\pi }}{V_0}} -
\frac {\bar q_z \sqrt {\pi}}{mn_0{V_0}^3} ,\
\\
\di V_1^-= \frac 12  {\frac { nU \sqrt {\pi }}{V_0 n_0}}- \frac 12
{\frac { P_{zz}}{mn_0{V_0}^2}}+ \frac 32 {\frac { k \rho T}{n_0
m^2{V_0}^2}}- \frac 12 {\frac {n}{n_0}}- \frac {q_z \sqrt {\pi}}{
mn_0{V_0}^3} - \frac 12 {\frac {U\sqrt {\pi }}{V_0}}+  \frac {
\bar q_z \sqrt {\pi}}{mn_0{V_0}^3} ,\
\\
\di U_1^+= {\frac {U}{V_0}} + \frac 32 \frac {\sqrt {\pi}
P_{zz}}{mn_0{V_0}^2} - \frac 32 \frac {\sqrt {\pi} k \rho
T}{n_0m^2{V_0}^2} + 8 \frac {\bar q_z}{mn_0{V_0}^3} - 4 \frac {
q_z}{mn_0{V_0}^3} ,\
\\
\di U_1^-={\frac {U}{V_0}}- \frac 32 {\frac {\sqrt {\pi}
P_{zz}}{mn_0 {V_0}^{2}}} + \frac 32 {\frac {\sqrt {\pi} k \rho
T}{n_0 m^2{V_0}^2}}+ 8{\frac { \bar q_z}{mn_0{V_0}^{3}}}-4{\frac {
q_z} {mn_0{V_0}^3}} .\
\end{array}
$$

 The values of integrals (\ref {q2}) as
functions of thermodynamic parameters of the system (\ref {q6})
are linked to the thermodynamic variables as:
\begin{equation}
\label{q10}
\begin{array}{l}
\di J_1=-\frac 52 \rho (\frac {kT_0}{m})^2+ \frac {11}{4} {\frac {
kT_0 P_{zz}}{m}}+ \frac 94 (\frac {k}{m})^2 \rho T_0 T ,\ \\
\di J_2= - \frac 32 \rho (\frac {kT_0}{m})^2 + \frac 94 {\frac {
kT_0  P_{zz}}{m}}+ \frac 34  (\frac {k}{m})^2 \rho T_0 T .\
\end{array}
\end{equation}

So we have closed the system (\ref {q6}), hence  a modification of
the procedure for deriving fluid mechanics (hydrody\-namic-type)
equations from the kinetic theory is proposed, it generalizes the
Navier-Stokes at arbitrary density (Knudsen numbers).

Our method gives a reasonable agreement with the experimental data
in the case of homogeneous gas \cite{LSV}. In the paper \cite{LSV}
the expressions for ($\di J_{1,2}$) are obtained with account some
nonlinear terms, that finally lead to more exact results.

%

 {\section{Stationary case (undisturbed atmosphere).}

Let's linearize the system (\ref{q6}) this way:
$$
\begin{array}{lll}
 \rho = \rho_0(z)(1+ \varepsilon \rho_1(t,z)) \ ,  &
 U_z = \varepsilon U_{z1}(t,z) \   , \vspace{2mm} \\
 P(t,z) = P_0(z)(1+ \varepsilon P_1(t,z)) \ ,  &
 T = T_0(z)(1 + \varepsilon T_1(t,z)) \ , \vspace{2mm} \\
 q_z = q_{z0}(z)(1 + \varepsilon q_{z1}(t,z)) \ , \quad &
 \bar q_z = \bar q_{z0}(z)(1 + \varepsilon \bar q_{z1}(t,z) ) \ , \quad &
  \quad \varepsilon << 1 \ .
\end{array}
$$

We obtain in the zero order:
\begin{equation}
\label{qwqw}
\begin{array}{l}
 \di
{\frac {{\frac {d}{dz}}P_{{0}}(z) }{\rho_{{0}}  (z) }}+g=0 \ ,\vspace{2mm} \\
 \di
 {\frac {d}{dz}}q_{0}(z)= 0 \ ,\vspace{2mm} \\
 \di
 2 {\frac {d}{dz}}\bar q_{z0}(z) +\nu(z) (P_{0}(z) -{\frac {k\rho_{0}(z) T_{0}(z)}{m}})
 = 0 \ ,\vspace{2mm} \\
\di -\frac 14 \frac {k^2}{m^2} T_{0}^2(z) { \frac
{d}{dz}}\rho_{0}(z) - \frac 12 \frac {k^2}{m^2} \rho_{0}(z)
T_{0}(z) {\frac {d}{dz}}T_{0}(z) + \nu(z) q_{0}(z) +
\ \\
\di + \frac 54 \frac km T_{0}(z) \frac {d}{dz}P_{0}(z) -{\frac
{P_{0}(z)}{\rho_{0}(z)}} \frac {d}{dz}P_{0}(z)+ \frac {11}{4}
\frac km P_{0}(z) {\frac {d}{dz}}T_{0}(z)=0 ,\
\vspace{2mm} \\
\di \nu (z) \bar q_{z0}(z) + \frac 94 \frac km P_{0}(z) {\frac
{d}{dz}}T_{0}(z) - \frac 32 \frac {k^2}{m^2} \rho_{0}(z)T_{0}(z)
{\frac {d}{dz}}T_{0}(z) - \
 \\
\di - \frac 34 \frac {k^2}{m^2} T_{0}^2(z) { \frac
{d}{dz}}\rho_{0}(z) + \frac 94 \frac km T_{0}(z) \frac
{d}{dz}P_{0}(z) - \frac 32 {\frac {P_{0}(z) {\frac
{d}{dz}}P_{0}(z)}{\rho_{0}(z)}}=0 .\
\end{array}
\end{equation}
Some version of such system that leads to a non-exponential
density dependence on height was studied in \cite{VerLe,LRVW}, the
paradox was discussed at \cite{RWV}.

Let's solve the zero order system.

$q_{z0}=C_1$. If $P_0= \frac km \rho_0 T_0$, then $\bar
q_{z0}=C_2= \frac 35 C_1$. If $ C_1=0 $, then $ T_0=C_3=const$ and
we'll have exponential density dependence on height.



We obtain in the first order:
\begin{equation}
\label{pq5}
\begin{array}{l}
\di
 \frac {\p }{\p t}\rho_1 + V_T \frac {\p}{\p z} U_{1} - \frac {V_T}{H} U_1
 = 0 \ ,\vspace{2mm} \\
 \di
 \frac {\p U_1}{\p t} + \frac{V_T}{2}\frac {\p P_1}{\p z}+
 \frac{V_T}{2H}(\rho_1-P_1)
= 0 \ ,\vspace{2mm} \\
 \di
\frac {\p \rho_1}{\p t} + \frac {\p T_1}{\p t}+\frac 53 V_T \frac
{\p U_1}{\p z} +\frac 23 V_T \frac {\p q_1}{\p z} -
\frac{V_T}{H}(U_1 + \frac 23 q_1 ) =0
 \ ,\vspace{2mm} \\
\di \frac {\p }{\p t}P_1 +3 V_T \frac {\p U_1}{\p z} +2 V_T \frac
{\p \bar q_1}{\p z} - \frac{V_T}{H}(2 \bar q_1 +U_1) + \nu
(P_1-\rho_1-T_1)=0
\ ,\vspace{2mm} \\
\di \frac {\p q_1}{\p t} + \frac 18 V_T \frac {\p P_1}{\p z} +
\frac 98 V_T \frac {\p T_1}{\p z} - \frac 18 V_T \frac {\p
\rho_1}{\p z} - \frac 38 \frac{V_T}{H}(\rho_1+T_1-P_1) + \nu q_1=0
\ , \vspace{2mm} \\
\di \frac {\p \bar q_1}{\p t} + \frac 38 V_T (\frac {\p P_1}{\p z}
+ \frac {\p T_1}{\p z}-\frac {\p \rho_1}{\p z}) + \frac 38 \frac
{V_T}{H} (P_1-\rho_1-T_1) +\nu \bar q_1=0 \ .
\end{array}
\end{equation}

{\section{Construction of solutions of the fluid dynamics system
by WKB method.}

In this section we apply the method WKB to the system~(\ref{pq5}).
We shall assume, that on the bottom boundary at $z=0$ a wave with
characteristic frequency $\omega_0$ is generated. Next we choose
the frequency $\omega_0$ to be large enough, to put characteristic
parameter $\xi=\di\frac{3\omega_0 H}{\vT} \gg 1$. We shall search
for the solution in the form:
\begin{equation}
\label{qr01} M_n=\psi_n \exp(i \omega_0 t)+c.c.\ ,
\end{equation}
where, for example, $\psi_1$, corresponding to the moment $M_1$,
is given by the expansion:
\begin{equation}
\label{qr1}
 \psi_1=\sum_{k=1}^6 \sum_{m=1}^\infty \di\frac{1}{(i \xi)^m} A_m^{(k)} \exp(i \xi \varphi_k(z))\ ,
\end{equation}
here $\varphi_k (z)$ - the phase functions corresponding to
different roots of dispersion relation. For other moments $M_n\,,
\quad n=2, \dots, 6$ corresponding functions $\psi_n$ are given by
similar to~(\ref{qr1}) expansion. The appropriate coefficients of
the series we shall designate by corresponding  $B_m^{(k)} \,
C_m^{(k)} \, D_m^{(k)} \, E_m^{(k)} \, F_m^{(k)}$. Substituting
the series~(\ref{qr1}) at the system~(\ref{pq5}) one arrives at
algebraic equations for the coefficients of  ~(\ref {qr1}) in each
 order. The condition
of solutions existence  results in the mentioned dispersion
relation:

\begin{equation}
\label{qr2}
\begin{array}{l}
 \frac {54}{125} {\eta}^{3} + \left( - \frac {12}{5} i u - \frac {63}{25} + \frac 35 u^2\right)
{\eta}^{2}+
 \left( -iu^3 + \frac {37}{5}iu - \frac {24}{5}u^2 +\frac {18}{5} \right) \eta - \vspace{2mm}\\
 -1-3iu +3u^2 +iu^3=0
\end{array}
\end{equation}

Here for convenience the following designations are introduced:
$$
 \left(\frac{\p \varphi_k}{\p
\bar z}\right)^2 = \di \frac{2}{15}\eta_k\ , \qquad
u=\frac{\nu_0}{\omega_0} \exp(-\bar z)\ ,
$$
where $\di \bar z = \frac {z}{H} $. For the coefficients
$A_1^{(k)} \, B_1^{(k)} \, \dots $ the algebraic relations are
obtained:

$$
\begin{array}{rcl}
      B_1^{(k)}&=& \mp \di \frac {\sqrt{30}}{6} \frac {A_1^{(k)}}{\sqrt{\eta_k}} \ ,\qquad
      C_1^{(k)} = \frac 13 {\frac { A_1^{(k)} ( -25+20iu+3\eta )}{-10+10iu+9\eta}}
 ,\qquad
      D_1^{(k)} = \frac 53 \frac {A_1^{(k)}}{\eta_k}  ,\vspace{2mm}\\
      E_1^{(k)} &=& \pm \di \frac {5}{12} {\frac { A_1^{(k)} \sqrt {30} ( 1+3 \eta) }{\sqrt {\eta} (-10+10iu+9\eta) }} \ ,
      \vspace{2mm}\\
      F_1^{(k)} &=& \pm \di \frac{1}{36} \sqrt{\frac {30}{\eta}} {\frac {  ( 50-100iu-135\eta-50{u}
^{2}+190iu\eta+81{\eta}^{2}+50{u}^{2}\eta-30i{\eta}^{2}u) }{ \eta
(-10+10iu+9\eta) }}A_1^{(k)}
 \
.\\
   \end{array}
$$

The dispersion relation~(\ref {qr2}) represents the cubic equation
with variable coefficients, therefore the exact analytical
solution by formula  Cardano looks very bulky and inconvenient for
 analysis. We study the behavior of solutions at $ \nu \to 0 $ (free
molecular regime) and
 $\nu\to \infty$ (a hydrodynamical regime).

At the limit of collisionless gas $\nu=0$ the dispersion relation
becomes:
$$
 \frac {54}{125} {\eta}^{3}   - \frac {63}{25}{\eta}^{2}+
 \frac {18}{5} \eta - 1=0\ .
$$
The roots are:
$$
 \eta_1 \approx 3.80\ ,\quad \eta_2 \approx 0.37\ ,\quad
 \eta_3 \approx 1.67\ .
$$
In a limit $\nu\to \infty $ (a hydrodynamical limit) for
specifying roots~(\ref {qr2}) by the theory of perturbations up to
$u^3$ for the three solutions branches  it is obtained:

$$
\begin{array}{l}
\smallskip
\eta_1=1.00 - 2.32 u^{-2} + i(1.20 u^{-1} - 4.88 u^{-3})\ ,\\
\smallskip
\eta_2= i 1.67 u  + 2.33 + 0.64 u^{-2} \ ,\\
\smallskip
\eta_3=- 1.39 u^2 + 2.50 + i(3.89 u - 1.20 u^{-1})\ .\\
\end{array}
$$
The first root relates to the acoustic branch. Accordingly, for
the $k_{i,\pm} =\pm \sqrt{\eta_i}$ we have:
$$
\begin{array}{l}
\smallskip
k_{1,+}\approx 1.00 - 0.98 u^{-2} + i(0.60 u^{-1} - 1.85 u^{-3})\ ,\\
\smallskip
k_{2,+}\approx\sqrt{u} (1-i)(0.64 u^{-1} + 0.019 u^{-3}) +
                   \sqrt{u} (1+i)(0.91  + 0.22 u^{-2})\ ,\\
\smallskip
k_{3,+}\approx 1.65 - 0.64 u^{-2} + i(1.18 u  + 0.094 u^{-1})\ .\\
\end{array}
$$

The solution of the equation~(\ref {qr2}) at any $u$ is evaluated
numerically. As an illustration let us consider a problem of
generation and propagation of a gas disturbance, by a plane
oscillating with a given frequency $\omega_0$. We restrict
ourselves by the case of homogeneous gas, because it is the only
case of existing experimental realization.  We evaluate
numerically the propagation velocity and attenuation factor of a
linear sound.

\includegraphics [width=14cm, height=10cm] {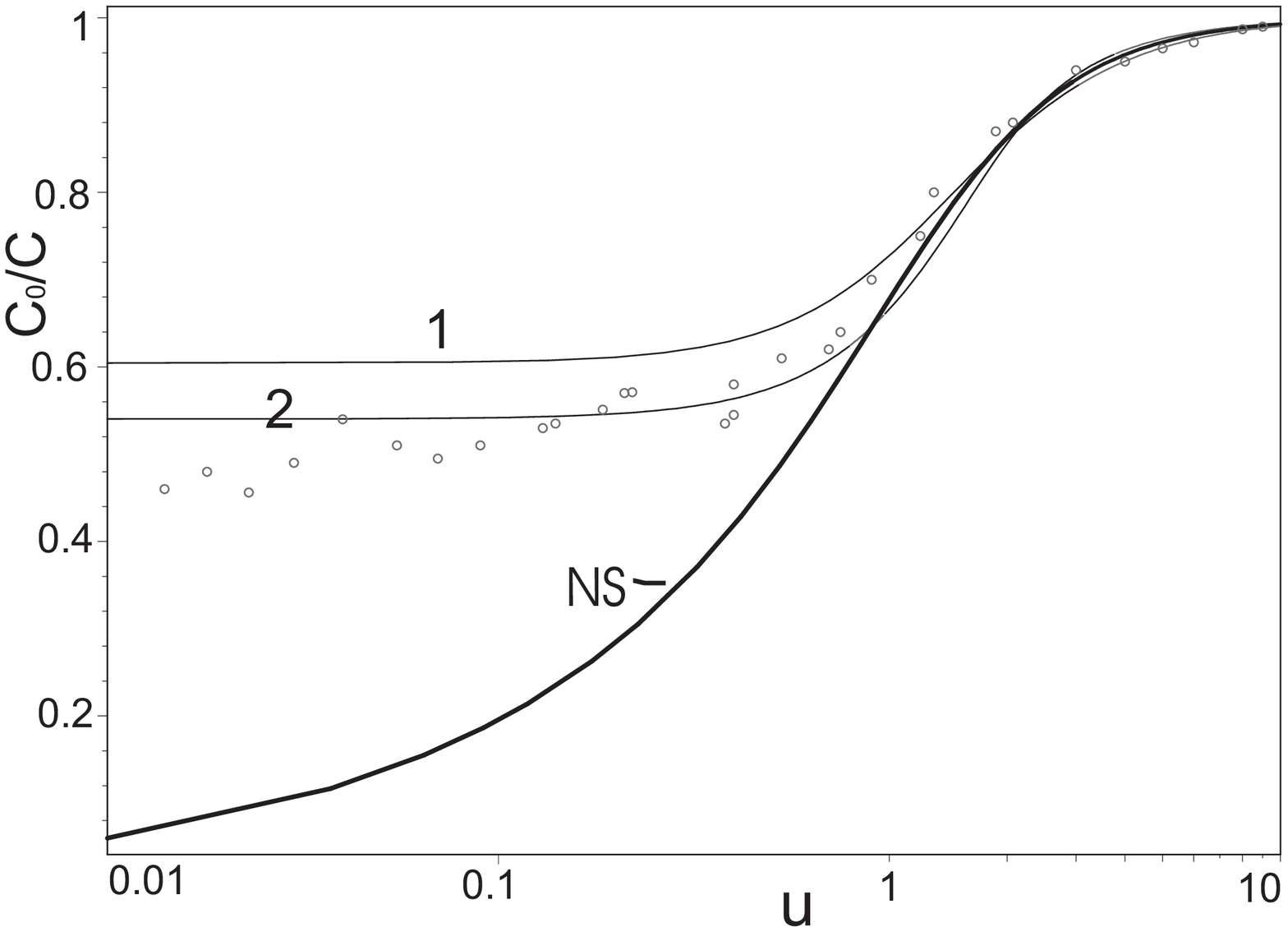}
\begin {center}
Fig. 1. The inverse non-dimensional phase velocity as a function
of the inverse Knudsen number. The results of this paper-1 are
compared to Navier-Stokes, previous our work \cite{LSV}-2 and the
experimental data of Meyer-Sessler \cite{meyer}-circle.
\end {center}

\includegraphics [width=14cm, height=10cm] {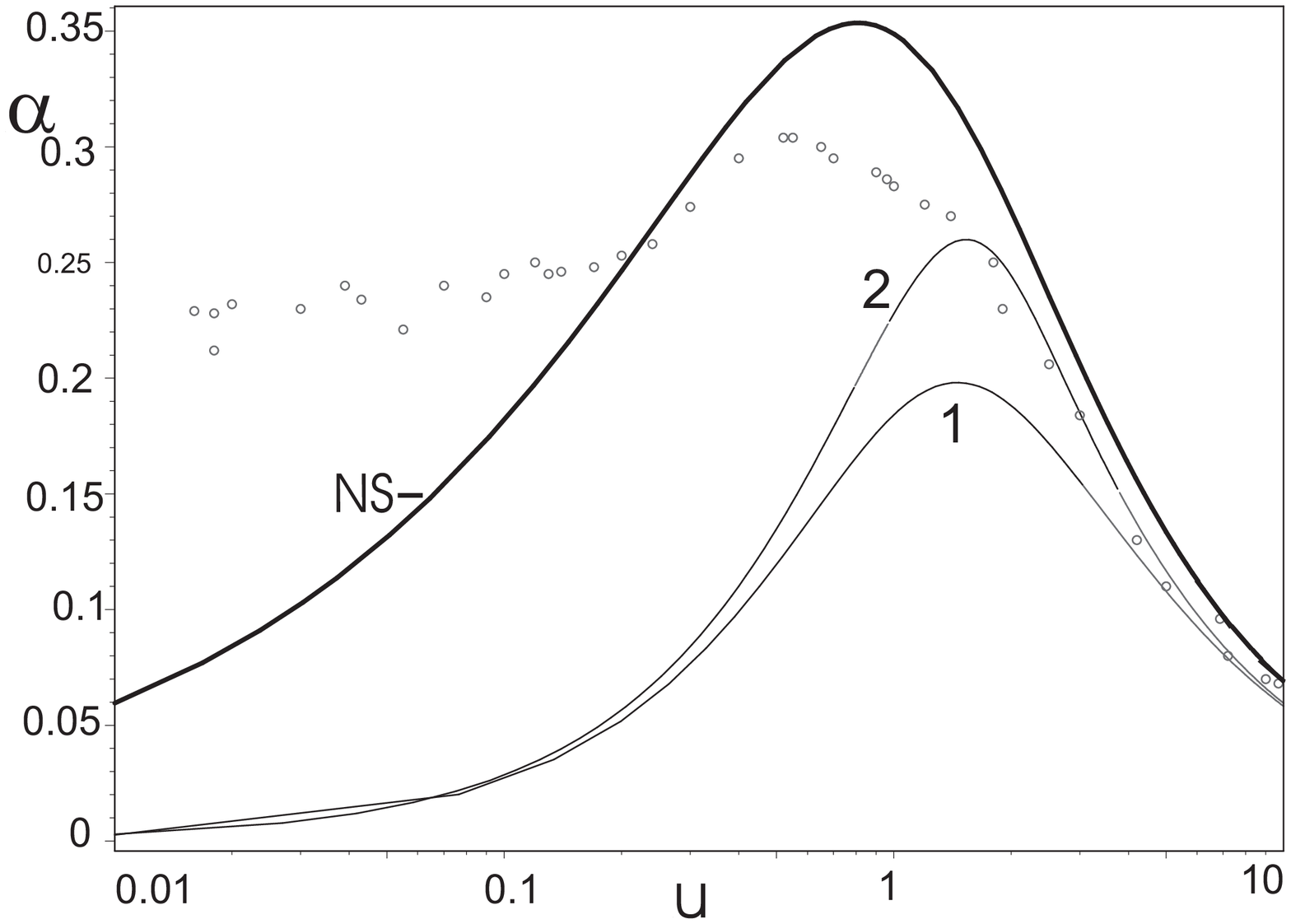}
\begin {center}
Fig. 2. The attenuation factor of the linear disturbance as a
function of the inverse Knudsen number.
\end {center}

\section{Conclusion}
In this paper we propose a one-dimensional theory of linear
disturbances in a gas, stratified in gravity field, hence
propagating through regions with crucially different Kn numbers.
The regime of the propagation dramatically changes from a
typically hydrodynamic  to the free-molecular one. We also studied
three-dimensional case \cite{Budapest2005}. Generally the theory
is based on Gross-Jackson kinetic equation, which solution is
built by means of locally equilibrium distribution function with
different local parameters for molecules moving "up" and "down".
Equations for six moments yields in the closed fluid mechanics
system. For the important generalizations of the foundation of
such theory see the recent review of Alexeev \cite{{alexeev2004}}.

\section{Acknowledgements}

We would like to thank
\begin{tabular}{|c|}
  \hline
   Vereshchagin D.A. \\
  \hline
\end{tabular} for important discussions.


\end {document}